\documentclass[a4paper]{article}
\usepackage{amssymb}
\newcommand{\bc}{\begin{center}}
\newcommand{\ec}{\end{center}}
\newcommand{\be}{\begin{equation}}
\newcommand{\ee}{\end{equation}}
\newcommand{\bea}{\begin{eqnarray}}
\newcommand{\eea}{\end{eqnarray}}
\newcommand{\ba}{\begin{array}}
\newcommand{\ea}{\end{array}}
\newcommand{\edc}{\end{document}}

\newcommand{\dsf}{\displaystyle\frac}

\def\w{\omega}
\def\O{\Omega}
\def\t{\theta}
\def\b{\beta}
\def\G{\Gamma}
\def\a{\alpha}

\def\s{\sigma}
\def\m{\mu}

\def\C{\mathbb{C}}
\def\R{\mathbb{R}}
\def\Z{\mathbb{Z}}
\def\Q{\mathbb{Q}}

\def\L{\Lambda}

\def\M{\cal M}

\def\i{{\bf 1}\!\!{\rm I}}
\begin{document}
\large
\begin {center}

{\Large {\bf ON GIBBS MEASURES OF MODELS WITH COMPETING TERNARY
AND BINARY  INTERACTIONS AND CORRESPONDING VON NEUMANN ALGEBRAS II
}\\[10mm] {\bf Farruh Mukhamedov}\footnote{CNR-Fellowship,
Dipartimento di Matematica, II Universita di Roma ("Tor Vergata"),
Via Della Ricerca Scientifica, Rome 00133, Italy, E-mail:
far75m@yandex.ru. On leave from Department of Mathematics,
National University of Uzbekistan, Tashkent, 700095, Uzbekistan.
}\ \ {\bf Utkir Rozikov}\footnote{Institute of Mathematics, 29, F.
Hodjaev str., Tashkent, 700143, Uzbekistan,
E-mail: rozikovu@yandex.ru}} \ec 

In the present paper the Ising model with competing binary ($J$)
and binary ($J_1$) interactions with spin values $\pm 1$, on a
Cayley tree of order 2 is considered. The structure of Gibbs
measures for the model considered is studied. We completely
describe the set of all periodic Gibbs measures for the model with
respect to any normal subgroup of finite index of  a group
representation of the Cayley tree. Types of von Neumann algebras,
generated by GNS-representation associated with diagonal states
corresponding to the translation invariant Gibbs measures, are
determined. It is proved that the factors associated with minimal
and maximal Gibbs states are isomorphic, and if they are of type
III$_\lambda$ then the factor associated with the unordered phase
of the model can be considered as a subfactors of these factors
respectively. Some concrete examples of factors are given too. \\[10mm]

\section{Introduction}

The present paper is  a continuation of the paper \cite{MR}. In
that paper it has been given motivations to study of the Ising
models with competing interactions and we have investigated the
Ising model with ternary interactions on a Cayley tree.

Recall that  the Cayley tree $\Gamma^k$ of order $ k\geq 1 $ is an
infinite tree, i.e., a graph without cycles, such that each vertex
of which lies on $ k+1 $ edges. Let $\Gamma^k=(V, \Lambda),$ where
$V$ is the set of vertices of $ \Gamma^k$, $\Lambda$ is the set of
edges of $ \Gamma^k$. The vertices $x$ and $y$ are called {\it
nearest neighbors},  if there exists an edge connecting them, such
vertices are denoted by $<x,y>$. The distance $d(x,y), x,y\in V$,
on the Cayley tree, is the length of the shortest path from $x$ to
$y$.

We set
$$ W_n=\{x\in V| d(x,x^0)=n\},  \ \ \ V_n=\cup_{m=1}^n W_m, $$
for an arbitrary point $ x^0 \in V $.

Denote
$$
S(x)=\{y\in W_{n+1} :  d(x,y)=1 \}, \ \ x\in W_n,
$$
this set is called a set of direct successors of $x$.

Two vertices $x,y\in V$ is called {\it one level
next-nearest-neighbor  vertices} if there is a vertex $z\in V$
such that  $x,y\in S(z)$, and they are denoted by $>x,y<$. In this
case the vertices $x,z,y$ was called {\it ternary} and denoted by
$<x,z,y>$.

In this paper we consider  the Ising model with competing
interactions, where the spin takes values in the set
$\Phi=\{-1,1\}$, on the Cayley tree  which is defined by the
following Hamiltonian
$$
H(\s)=-J \sum\limits_{>x,y<}{\s(x)\s(y)} -J_1
\sum\limits_{<x,y>}{\s(x)\s(y)}  \eqno (1.1)
$$
where $J,J_1\in {\R}$ are coupling constants and $\s$ a
configuration on $V$, i.e. $\s\in\Omega=\Phi^{V}$.

The other parts of the paper is organized as follows. In section 2
using the similar argument as \cite{MR}  we reduce the problem of
describing limit Gibbs measures to the problem of solving a
nonlinear functional equation. By means of the  obtained equation
we construct periodic Gibbs measures and find ground states the
model. In section 3, we determine the types of von Neumann
algebras generated by GNS-representation associated with diagonal
states corresponding to the translation invariant measures. In
addition, we will demonstrate more concrete examples of factors.
In section 4 we discuss the results.
\\[2mm]

\section{On the set of Gibbs measures}

In this section recall the construction of a special class of
limiting Gibbs measures for the  Ising model on a Cayley tree with
competing interactions.

Let $h:x\to {\R}$ be a real
valued function of $x\in V$. Given $n=1,2,...$ consider the probability
measure $\m^{(n)}$ on $\Phi^{V_n}$ defined by
$$
\mu^{(n)}(\s_n)=Z^{-1}_{n}\exp\{-\b H(\s_n)+\sum_{x\in
W_n}h_x\s(x)\}, \eqno(2.1)
$$
Here,  $\b=\frac{1}{T}$ and  $\s_n:x\in V_n\to\s_n(x)$ and $Z_n$
is the corresponding partition function:
$$
Z_n=\sum_{\tilde\s_n\in\Omega_{V_n}}\exp\{-\b
H(\tilde\s_n)+\sum_{x\in W_n}h_x\tilde\s(x)\},
$$
$$ H(\s_n)=-J
\sum\limits_{>x,y<: x,y\in V_n}{\s_n(x)\s_n(y)} -J_1
\sum\limits_{<x,y>: x,y\in V_n}{\s_n(x)\s_n(y)} \eqno(2.2)$$

Recall that the consistency condition for $\m^{(n)}(\s_n), n\geq
1$ is
$$
\sum_{\s^{(n)}}\m^{(n)}(\s_{n-1},\s^{(n)})=\m^{(n-1)}(\s_{n-1}),
\eqno(2.3)
$$
where $\s^{(n)}=\{\s(x), x\in W_n\}$.

The following statement describes conditions on $h_x$ guaranteeing
the consistency condition of measures $\m^{(n)}(\s_n)$. In the
sequel for the simplicity we will consider the case $k=2$.

{\bf Theorem 2.1.} {\it The measures  $\m^{(n)}(\s_n)$,
$n=1,2,...$ satisfy the consistency condition (2.3) if and only if
for any $x\in V$ the following equation holds: $$ h_x= {1 \over
2}\log\bigg(\frac{\theta_1^2\theta
e^{2(h_y+h_z)}+\theta_1(e^{2h_y}+ e^{2h_z})+\theta}{\theta
e^{2(h_y+h_z)}+\theta_1(e^{2h_y}+
e^{2h_z})+\theta_1^2\theta}\bigg) \eqno (2.4) $$ here
$\theta=e^{2\beta J}, \ \ \theta_1=e^{2\beta J_1}$ and $<y,x,z>$
are ternary neighbors}.

{\bf Proof.} Using (2.1) it is easy to see that (2.3) and (2.4)
are equivalent.(cf. \cite{MR}).\\[3mm]

This theorem reduces the problem of describing of Gibbs measures
to the description of solutions of the functional equation (2.4).

According to Proposition 2.1\cite{MR} that there exists a
one-to-one correspondence between the set  $V$ of vertices of the
Cayley tree of order $k\geq 1$ and the group $G_{k}$ of the free
products of $k+1$ cyclic  groups  of the second order with
generators $a_1,a_2,...,a_{k+1}$.

Recall that $h=\{h_x : x\in G_k\}$ is {\it $\hat G_k$-periodic} if
$h_{yx}=h_x$ for all $x\in G_k$ and $y\in\hat G_k$, here $\hat
G_k$ is a normal subgroup of $G_k$ with finite index. A Gibbs
measure is called {\it $\hat G_k$-periodic} if it corresponds to
$\hat G_k$-periodic function $h$. If it is $G_k$-periodic, then
this measure is {\it translation-invariant}.

As before in the sequel we will consider the group $G_2$.

As in \cite{MR} we firstly find translation - invariant solutions
of (2.4). This case recently has been investigated in
Ref.\cite{GPW}. For the sake of completeness and since throughout
the paper we will use this result, we recall it.

In this setting (2.4) has the form
$$
u={\theta_1^2\theta u^2+2\theta_1 u+ \theta \over \theta
u^2+2\theta_1 u+\theta_1^2\theta } \eqno  (2.5)$$ where
$u=e^{2h}$.

\indent{\bf Proposition 2.2.} \cite{GPW}.{\it  If $\theta_1
>\sqrt{3}$ and $\theta>\dsf{2\theta_1}{\theta_1^2-3}$ then  for
all pairs $(\theta,\theta_1)$ the equation (2.5) has three
positive solutions $u^*_{1}<u_{2}^{*}<u_{3}^*$, here $u^*_2=1$.
Otherwise (2.5) has a unique solution $u_*=1$.}

{\bf Remark 2.1.} The numbers $u^*_{1}$ and $u_{3}^*$ are the
solutions of the following equation
$$
u^2+(1+\a)u+1=0, \eqno(2.6)
$$
here $\a=\dsf{2\t_1}{\t}-\t_1^2$. Hence, if $\b\to\infty$ then
$u^*_3\to\infty$ and $u^*_1\to 0$.

By $\mu_1,\mu_2,\mu_3$  we denote Gibbs measures corresponding to
these solutions.

Using the same argument as \cite{MR} one can prove the following

{\bf Theorem 2.3.} {\it For the model (1.1) with parameters
$J_1>0$ and $J\in {\R}$ on the Cayley tree $\G^2$ the following
assertions hold}
\begin{enumerate}
   \item[(i)]{\it if $\theta_1 >\sqrt{3}$,
$\theta>\dsf{2\theta_1}{\theta_1^2-3}$ then the measures $\m_1$ and $\m_3$ are extreme;}
   \item[(ii)] {\it  in the opposite case there is a  Gibbs measure
$\m_*(=\m_2)$ and it is extreme.}
\end{enumerate}

{\bf Remark 2.2.} This theorem specifies the result obtained in
\cite{GPW}, as there it  was proved that a phase transition occurs
if and only if the above indicated conditions is satisfied, and
the extremity was open.  The formulated theorem answers that the
found Gibbs measures are extreme. In spite of this, further we
will show that a phase transition can be occur when the condition
of theorem is not satisfied.

{\bf Remark 2.3.} The measure $\m_2$ corresponding
 to the solution $h_x=0$, $x\in V$ is the unordered phase, i.e. the spin $\s(x)$ takes
its values $\pm 1$ with respect to $\m_2$ with probability $1/2$.\\

Now we turn to the constructions of periodic Gibbs measures. Let
$H_0$ be a subgroup of index $r$ in $G_2$, and let
$G_2|H_0=\{H_0,H_1,...,H_{r-1}\}$ be the quotient group. Let
$g_i(x)=|S^*(x)\cap H_i|, i=0,1,...,r-1$; $N(x)=|\{j:q_j(x)\ne
0\}|,$ where $S^*(x)$ is the set of all nearest neighbors of $x\in
G_2.$ Denote $Q(x)=(q_0(x),q_1(x),...,q_{r-1}).$ We note (see
\cite{R}) that for every $x\in G_2$ there is a permutation $\pi_x$
of the coordinates of the vector $Q(e)$ (where $e$ is the identity
of $G_2$) such that
$$ \pi_xQ(e)=Q(x). \eqno(2.7)
$$
It follows from (2.7) that $N(x)=N(e)$ for all $x\in G_2.$

It is clear that each $H_0-$ periodic function $h_x$  is given by
$$\{h_x=h_i \ \ \mbox{for} \ \ x\in H_i, i=0,1,...,r-1\}.$$

By $G^{(2)}_2$ we denoted in \cite{MR} the subgroup of $G_2$
consisting of all words with even length. This $G^{(2)}_2$ has an
index 2.

{\bf Theorem 2.4.} {\it Let $H_0$ be a subgroup of finite index in
$G_2$. Then each $H_0-$ periodic Gibbs measure for (1.1) model is
either translation-invariant or $G^{(2)}_2-$ periodic.}

{\bf Proof.} Let $f(x,y)$ be function defined as follows
$$
f(x,y)=\frac{\theta_1^2\theta xy+\theta_1(x+y)+\theta} {\theta
xy+\theta_1(x+y)+\theta_1^2\theta},  \ \  x,y>0. \eqno(2.8)
$$
For $\theta_1\ne 1$ it is easy to see that $f(u_1,v)=f(u_2,v)$ if
and only if $u_1=u_2$. Also
 $f(u,v_1)=f(u,v_1)$ if and only if $v_1=v_2.$ Using this   property of $f(u,v)$, by Theorem 2.1 and (2.7) we have
$$ h_x=h_y=h_1, \ \ \mbox{if} \ \  x,y\in S^*(z), z\in G^{(2)}_2;$$
$$ h_x=h_y=h_2, \ \ \mbox{if} \ \  x,y\in S^*(z), z\in G_2\setminus G^{(2)}_2.$$
Thus the measures are translational-invariant (if $h_1=h_2$) or
$G^{(2)}_2-$ periodic (if $h_1\ne h_2$). The theorem is proved.

If $H_0$ is  a  subgroup of finite index in $G_2$, then it natural
to ask: what condition on $H_0$ guarantees that each
$H_0-$periodic Gibbs measure to be translation-invariant? We put
$I(H_0)=H_0\cap \{a_1,a_2,a_3\},$ where $a_i, i=1,2,3$ are
generators of $G_2$.

{\bf Theorem 2.5.} {\it If $I(H_0)\ne \emptyset$, then each $H_0-$
periodic Gibbs measure for (1.1) model is translation-invariant.}

{\bf Proof.} Take $x\in H_0.$ We recall that the inclusion
$xa_i\in H_0$ holds if and only if $a_i\in H_0.$ Since $I(H_0)\ne
\emptyset$, there is an element $a_i\in H_0.$ Therefore $H_0$
contains the subset $H_0a_i=\{xa_i: x\in H_0\}$. By Theorem 2.4 we
have $h_x=h_1$ and $h_{xa_i}=h_2.$ Since $x$ and $xa_i$ belong to
$H_0$, it follows that $h_x=h_{xa_i}=h_1=h_2.$ Thus each $H_0-$
periodic Gibbs measure is translation -invariant. This proves the
theorem.

Theorems 2.4, 2.5 reduce the problem of describing $H_0-$ periodic
Gibbs measures with $I(H_0)\ne \emptyset$ to describing the fixed
points of the function $f(u,u)$ (see (2.8)) which describe the
translation-invariant Gibbs measures. If $I(H_0)=\emptyset$, this
problem is reduced to the describing solutions of the system:

$$ \left\{ \ba{ll}
u=\dsf{\theta_1^2\theta v^2+2\theta_1v+\theta}
{\theta v^2+2\theta_1 v+\theta_1^2\theta }, \\[2mm]
v=\dsf{\theta_1^2\theta u^2+2\theta_1u+\theta} {\theta
u^2+2\theta_1u+\theta_1^2\theta }, \\ \ea \right. \eqno  (2.9)
$$ where $u=\exp\{2h_1\}, v=\exp\{2h_2\}$.

The analysis of the equation (2.9) is carried in the following

{\bf Proposition 2.6.} {\it The equation (2.9) has three positive
solutions $(1,1)$,$(u_*,v_*)$ and $(v_*,u_*)$ (here $u_*<v_*$) if
and only if $\theta_1<1/\sqrt{3}$ and
$\theta>\dsf{2\theta_1}{1-3\theta^2_1}$. Here $u_*,v_*$ are the
solutions of the equation:
$$
\t_1^2\t(\t_1^2\t+2\t_1+\t)(x^2+1)+((\t_1^2\t)^2+4\t_1^3\t+4\t_1^2-\t^2)x=0.
\eqno(2.10) $$ }

{\bf Proof.} It is clear that $(1.1)$ is a solution of $(2.9)$.
The equation (2.9) can be written as $u=g(g(u))$, here
$g(u)=f(u,u)$. Hence, the solutions of the equations $u=g(u)$ are
the solution of (2.9), but they describe only the translation -
invariant Gibbs measures.  Now we should find solutions of
$\dsf{g(g(u))-u}{g(u)-u}=0$. After some calculations it can be
shown that the last equation has the form (2.10).

Full analysis of the equation (2.10) shows that parameters
$\t,\t_1$ must satisfy the condition of the proposition. This
completes the proof.

Thus we can formulate the following

{\bf Theorem 2.7.} {\it For the model (1.1) with respect to any
subgroup $H_0$ of finite index the following assertions hold:}
\begin{enumerate}
   \item[(i)] {\it Let  $\t_1>\sqrt{3}$,  $\t>\dsf{2\t_1}{\t_1^2-3}$ be satisfied,
  then $H_0$-periodic Gibbs measures coincide with the translation - invariant Gibbs measures.}
  \item[(ii)] {\it Let   $\theta_1<1/\sqrt{3}$, $\theta>\dsf{2\theta_1}{1-3\theta^2_1}$ be satisfied.}
  \begin{enumerate}
  \item  {\it If $I(H_0)\neq \emptyset$ then $H_0$-periodic Gibbs measures
   coincide with the translation - invariant Gibbs measures.}
  \item {\it  If $I(H_0)=\emptyset$ then there are three $H_0$-periodic
  ($=G^{(2)}_2-$ periodic) Gibbs measures $\m_{12},\m_{21}$ and $\m_*$. Here the measure $\m_*$ corresponds
to the unique solution of equation (2.5).}
  \end{enumerate}
\end{enumerate}

{\bf Proof.} Let the condition (i) be satisfied. Then Proposition
2.6 implies that there is no $G^{(2)}_2$-periodic Gibbs measures
in this setting. Therefore, according to Theorem 2.4  we conclude
that that the $H_0$-periodic Gibbs measures are
translation-invariant. Now let (ii) hold. Then the assertions (a)
and (b) immediately follow from Proposition 2.6, Theorems 2.4. and
2.5. This completes the proof.

{\bf Remark 2.4.} In \cite{MR} we have investigated only
$G_2^{(2)}$-periodic Gibbs measures, the proved Theorem 2.7
completely describes all periodic Gibbs measures, associated with
subgroups of $G_2$ with finite index, of the model.\\

Now comparing Theorems 2.3 and 2.7 we infer the following

{\bf Corollary 2.8.} {\it If $1/\sqrt{3}<\t_1<\sqrt{3}$ then for
the model (1.1) there is no
phase transition.}\\

By using the similar argument as in \cite{MR} we can prove the
extremity of $\m_{12},\m_{21}$.

{\bf Theorem 2.9.} {\it  Let $\theta_1<1/\sqrt{3}$ and
$\theta>\dsf{2\theta_1}{1-3\theta^2_1}$ be satisfied. Then the
measures $\m_{12},\m_{21}$ and $\m_*$ are extreme.}\\

Using the analogical way as in \cite{BG},\cite{GR}  with the aid
of measures $\m_1,\m_2$ and $\m_3$  one can construct uncountable
number of extreme Gibbs measures.\\

From the construction of the Gibbs measures we easily see that the
measures $\m_1$ and $\m_3$ depend on parameter $\b$. Now we are
interested on the behaviour of these measures when $\b$ goes to
$\infty$.

Put
$$
\s_+=\{\s(x): \s(x)=1, \ \forall x\in\G^2\},
$$
$$
\s_-=\{\s(x): \s(x)=-1, \ \forall x\in\G^2\}.
$$

{\bf Theorem 2.10.} {\it Let  $\theta_1 >\sqrt{3}$ and
$\theta>\dsf{2\theta_1}{\theta_1^2-3}$, then
$$
\m_1\to \delta_{\s_-}, \ \ \ \m_3\to \delta_{\s_+} \ \ \ \textrm{as} \ \ \b\to\infty,
$$
here $\delta_\s$ is a delta-measure concentrated on $\s$.}

{\bf Proof.} Consider the measure $\m_3$. This measure corresponds
to the function $h_x=h_3,$ $x\in V$, here $h_3>0$ (see Proposition
2.2). Let us first consider a case:
$$
\m_3(\s(x)=1)=\dsf{e^{h_3}}{e^{h_3}+e^{-h_3}}=\dsf{u^*_3}{u^*_3+1}\to 1 \ \ \textrm{as} \  \b\to\infty,
$$
since $u^*_3\to\infty$ as $\b\to\infty$, here $x\in V$. Let us turn to the general case.
From the condition imposed in the Theorem we find that $J_1>0$. Now separately consider two cases.

{\tt First case.} Let $J>0$. Then from the form of Hamiltonian
(1.1) it is easy to check that $H(\s_n|_{V_n})\geq
H(\s_{+}|_{V_n})$ for all $\s\in \Omega$  and $n>0$. It follows
that \bea \m_3(\s_+|_{V_n})&=&\dsf{\exp\{-\b
H(\s_+|_{V_n})+h_3|W_n|\}}{\sum\limits_{\tilde\s_n\in\O_{V_n}}
\exp\{-\b H(\tilde\s_n)+h_3\sum\limits_{x\in W_n}\tilde\s(x)\}}\nonumber \\
&=&\dsf{1}{1+\sum\limits_{\tilde\s_n\in\O_{V_n},\tilde \s_n\neq
\s_+|_{V_n}}
\dsf{\exp\{-\b H(\tilde\s_n)+h_3\sum\limits_{x\in W_n}\tilde\s(x)\}}{\exp\{-\b H(\s_+|_{V_n})+h_3|W_n|\}}}\nonumber \\
&\geq&\dsf{1}{1+1/u^*_3}\to 1 \ \ \textrm{as} \
\b\to\infty.\nonumber \eea

The last inequality yields that $\m_3\to\delta_{\s_+}$.

{\tt Second case.} Let $J<0$. Let us introduce some notations.
\bea &&A(\s_n)=\sum_{>x,y<: x,y\in V_n}\s(x)\s(y), \ \
A=A(\s_+|_{V_n}), \nonumber \\
&& B(\s_n)=\sum_{<x,y>: x,y\in V_n}\s(x)\s(y), \ \
B=B(\s_+|_{V_n}), \nonumber\\
&& C(\s_n)=\sum_{x\in V_n}\s(x), \ \ C=C(\s_+|_{V_n}).\nonumber
\eea
Then it is easy to see that the following equality holds
$$
\m_3(\s_+|_{V_n})
=\dsf{1}{1+\sum\limits_{\tilde\s_n\in\O_{V_n},\tilde \s_n\neq \s_+|_{V_n}}
\dsf{1}{e^{J\b(A-A(\tilde\s_n))}e^{J_1\b(B-B(\tilde\s_n))}e^{h_3(C-C(\tilde\s_n))}}}.
$$
We want to show that
$$
\sum_{\tilde\s_n\in\O_{V_n},\tilde \s_n\neq \s_+|_{V_n}}
\dsf{1}{e^{J\b(A-A(\tilde\s_n))}e^{J_1\b(B-B(\tilde\s_n))}e^{h_3(C-C(\tilde\s_n))}}\to 0 \ \textrm{as}
 \ \b\to\infty.
$$
It is enough to prove that
$$
\dsf{1}{e^{J\b(A-A(\tilde\s_n))}e^{J_1\b(B-B(\tilde\s_n))}e^{h_3(C-C(\tilde\s_n))}}\to 0 \ \ \textrm{as}
 \ \ \b\to\infty
$$
for all $\tilde\s_n\in\O_{V_n},\tilde \s_n\neq \s_+|_{V_n}$. We rewrite the last sentence as
follows
$$
\dsf{1}{e^{J\b(A-A(\tilde\s_n))}e^{J_1\b(B-B(\tilde\s_n))}e^{h_3(C-C(\tilde\s_n))}}=
\dsf{1}{\t^{(A-A(\tilde\s_n))/2}\t_1^{(B-B(\tilde\s_n))/2}(u^*_3)^{(C-C(\tilde\s_n))/2}}\leq
$$
$$
\leq\dsf{(\t_1^2-3)^{(A-A(\tilde\s_n))/2}}{\t_1^{(A-A(\tilde\s_n)+B-B(\tilde\s_n))/2}
(u^*_3)^{(C-C(\tilde\s_n))/2}}\leq
$$
$$
\leq\dsf{(\t_1^2-3)^{(A-A(\tilde\s_n))/2}}{\t_1^{(A-A(\tilde\s_n)+B-B(\tilde\s_n))/2}
u^*_3}.\eqno(2.11)
$$
here we have used the inequality $\t>\dsf{2\t_1}{\t_1^2-3}$.

Obviously, if $\b$ is large enough we have \bea
\dsf{(\t_1^2-3)^{(A-A(\tilde\s_n))}}{\t_1^{A-A(\tilde\s_n)+B-B(\tilde\s_n)}}&\sim&
\dsf{\t_1^{2(A-A(\tilde\s_n))}}{\t_1^{A-A(\tilde\s_n)+B-B(\tilde\s_n)}}\nonumber\\
&=&\dsf{\t_1^{B(\tilde\s_n)-A(\tilde\s_n)}}{\t_1^{B-A}}. \nonumber
\eea

If $B(\tilde\s_n)-A(\tilde\s_n)\leq B-A$ then the last relation
implies that
$\dsf{(\t_1^2-3)^{(A-A(\tilde\s_n))}}{\t_1^{A-A(\tilde\s_n)+B-B(\tilde\s_n)}}
$ is bounded, and hence from (2.11) we get the required relation.

Now it remains to prove the following

{\bf Lemma 2.11.} {\it For every $n>0$ and $\s_n\in \O_{V_n}$ the
following inequality holds
$$
B(\s_n)-A(\s_n)\leq B-A.\eqno(2.12)
$$
}

{\bf Proof.} Denote ${\cal C}(\s_n)=\{x\in V_n : \s(x)=-1\}$.
Maximal connected components of ${\cal C}(\s_n)$ we will denote by
${\cal K}_1(\s_n),\cdots,{\cal K}_m(\s_n)$. For a connected subset
${\cal K}$ of $V_n$ put \bea &&\partial{\cal K}=\{x\in
V_n\setminus{\cal K} : \ <x,y> \ \textrm{for some}
 \ y\in{\cal K}\}, \nonumber \\
&& \partial^2{\cal K}=\{x\in V_n\setminus{\cal K} : \ >x,y< \
\textrm{for some} \ y\in{\cal K}\}.\nonumber \eea From definition
of $A(\s_n)$ and $B(\s_n)$ we get \bea
&& B(\s_n)=B-2\sum_j|\partial{\cal K}_j(\s_n)|, \nonumber \\
&& A(\s_n)=A-2\sum_j|\partial^2{\cal K}_j(\s_n)\setminus
\cup_{m\ne j}{\cal K}_m(\s_n)|,\nonumber \eea here $|A|$ stands
for a number of elements of a set $A$.

To prove (2.12) it enough to show that $|\partial^2{\cal K}|\leq
|\partial{\cal K}|$ for all connected subsets   ${\cal K}$ of
$V_n$. For each $x\in \partial^2 {\cal K}$ we can show a
$y=y(x)\in \partial {\cal K}$. Indeed, if $>x,t<, \ \ t\in {\cal
K}$ and $<x,z,t>$ then $y(x)=x $ if $z\in {\cal K}$ and $y(x)=z$
if
 $z\notin {\cal K}.$ It is clear $y(x)\in \partial {\cal K}.$ Now we will prove that
 if $x_1\ne x_2\in \partial^2 {\cal K}$ then $y(x_1)\ne y(x_2)\in \partial {\cal K}$.
 Let $<x_1,z_1,t_1>, <x_2,z_2,t_2>,$ where $t_1, t_2\in {\cal K}$. By definition of
 $y(x)$ we have $y(x_i)\in \{x_i,z_i\}, i=1,2$. So to prove $y(x_1)\ne
y(x_2)$ for $x_1\ne x_2$ it is enough to show $z_1\ne z_2$. Note that in our case (i.e. $k=2$)
if $x_1\ne x_2\in \partial^2 {\cal K}$ then $d(x_1,x_2)\geq 3$, since $<x_i,z_i>, i=1,2$ and
hence we get $z_1\ne z_2$. Thus
 $|\partial^2{\cal K}|\leq |\partial{\cal K}|.$
 So Lemma is proved.

By similar argument the theorem can be proved for the measure $\m_1$. Thus
the theorem is proved.

From Theorem 2.10 we conclude that $\s_+$ and $\s_-$ are ground
states of the considered model.

{\bf Remark 2.6.} If in the condition  of Theorem 2.10, we put
$J=0$ then the obtained result coincides with Theorem 2.3 of
\cite{BRZ2}.\\

Now introduce  two configurations as follows
$$
\s_{+-}=\{\s_{+-}(x), \ x\in V\}, \ \ \ \s_{-+}=\{\s_{-+}(x), \ x\in V\},
$$
where
$$
\s_{+-}(x)=
\left\{
\ba{ll}
1, \ \ \textrm{if} \ \ x\in G^{(2)}_2,\\
-1, \ \ \textrm{if} \ \ x\in G_2\setminus G_2^{(2)}, \ea \right. \
\ \s_{-+}(x)= \left\{ \ba{ll}
-1, \ \ \textrm{if} \ \ x\in G^{(2)}_2,\\
1, \ \ \textrm{if} \ \ x\in G_2\setminus G^{(2)}_2. \ea \right.
$$

We can formulate the following

{\bf Theorem 2.12.} {\it Let  $\theta_1<1/\sqrt{3}$ and
$\theta>\dsf{2\theta_1}{1-3\theta^2_1}$ then
$$
\m_{12}\to \delta_{\s_{-+}}, \ \ \ \m_{21}\to \delta_{\s_{+-}} \ \ \ \textrm{as} \ \ \b\to\infty.
$$
}

The proof is  similar to the proof of Theorem 2.10.\\

\section{Diagonal states generated by Gibbs measures and corresponding
von Neumann algebras}

In this section we consider a case $\theta_1 >\sqrt{3}$,
$\theta>\dsf{2\theta_1}{\theta_1^2-3}$  and determine types of von Neumann algebras generated
by the GNS - representation associated with the diagonal states corresponding
to the translation invariant measures.

As the paper \cite{MR} we consider  $C^*$-algebra
$A=\otimes_{{\G}^k}M_{2}(\C)$, where $M_{2}(\C)$ is the algebra of
$2{\rm x}2$ matrices over the field $\C$ of complex numbers.

By $\w_i$ ($i=1,2,3$) we denote the diagonal state generated by the translation invariant
measures $\mu_1,\m_2,\m_3$ respectively. On the finite dimensional $C^*$-subalgebra
$A_{V_n}=\otimes_{V_n}M_{2}(\C)\subset A$ we rewrite the state
$\w_i$ as follows
$$
\w_i(x)=\frac{tr(e^{\tilde H_i(V_n)}x)}{tr(e^{\tilde H_i(V_n)})},
\ \ x\in A_{V_n}, \eqno(3.1)
$$
where  $tr$ is a trace on $A_{V_n}$. The term $\s(x)\s(y)$,
($>x,y<$) in (2.2) we represent as a diagonal element of
$M_2(\C)\otimes M_2(\C)\otimes M_2(\C)$ in the standard basis as
follows
$$
\s(x)\s(y)=
\left (
\ba{cccccccc}
1&0&0&0&0&0&0&0 \\
0&-1&0&0&0&0&0&0 \\
0&0&1&0&0&0&0&0 \\
0&0&0&-1&0&0&0&0 \\
0&0&0&0&-1&0&0&0 \\
0&0&0&0&0&1&0&0 \\
0&0&0&0&0&0&-1&0 \\
0&0&0&0&0&0&0&1 \\
\ea \right ). \eqno(3.2)
$$
Using (3.2), the form of Hamiltonian (1.1),(2.1) and (3.1) the
Hamiltonian $\tilde H_i(V_n)$ in the standard basis of $A_{V_n}$
has the form
$$
\tilde H_i(V_n)=\sum_{>x,y<:x,y\in  V_n}F_{>x,y<}+\sum_{<x,y>:x,y\in  V_n}G_{<x,y>}+
\sum_{x\in W_{n}}h_i\s_x^z , $$
here and below
$$
F_{>x,y<}=
\left(
\begin{array}{cc}
A\otimes\i & O   \\
O  & B\otimes\i  \\
\end{array}
\right), \ \
A=
\left(
\begin{array}{cc}
\log p_1  & 0  \\
0  &  \log p_2  \\
\end{array}
\right), \ \
B=UAU, \ \
U=
\left(
\begin{array}{cc}
0 & 1  \\
1  & 0 \\
\end{array}
\right), \eqno (3.3)
$$
$$
p_1=\frac{1}{e^{-2\b J}+1}, \ \ p_2=1-p_1=\frac{e^{-2bJ}}{e^{-2\b
J}+1} \eqno(3.4)
$$
$$ G_{<x,y>}=
\left(
\begin{array}{cc}
A_1 & O  \\
O  & UA_1U \\
\end{array}
\right), \ \
A_1=
\left(
\begin{array}{cc}
\log p_{11}  & 0  \\
0  &  \log p_{22}  \\
\end{array}
\right), \ \ \eqno(3.5)
$$
$$
p_{11}=\frac{1}{e^{-2\b J_1}+1}, \ \ p_{22}=1-p_{11}=\frac{e^{-2bJ_1}}{e^{-2\b J_1}+1} \ \
\s_x^{z}=
\left(
\begin{array}{cc}
1 & 0  \\
0  & -1\\
\end{array}
\right), \ \ \eqno(3.6)
$$
and $h_i=\log u^*_i/2$, where $u^*_i$  is a solution of (2.5).

Hence the state $\w_i$ is an Gibbs state for quantized
Hamiltonian $$ \tilde H =
\sum_{>x,y<}F_{<x,y,z>}+\sum_{<x,y>}G_{<x,y>}.
$$

Denote ${\M}_i=\pi_{\w_i}(A)'',$ where $\pi_{\w_i}-$ is a GNS -
representation associated with $\w_i$ (see Ref. [BR1, definition
2.3.18]).  Note ${\M}_i$ is a factor, since the measures $\m_i$ ($i=1,2,3$) are translation invariant
and satisfy mixing property, i.e.
$$
\lim_{|g|\to\infty}\w_i(T_g(x)y)=\w_i(x)\w_i(y),
$$
here $T_g$ is a left shift transformation of $G_2$.  Our goal in
the present section is to determine a type of ${\M}_i$.

We note that the modular group of ${\M}_i$ associated with $\w_i$
is defined by $$ \s^{\w_i}_{t}(x)=\lim_{V_n\to V}\exp\{it\tilde
H_i(V_n)\}x\exp\{-it\tilde H_i(V_n)\}, \ \ x\in {\M}_i. \eqno(3.7)
$$ here as before
$$
\tilde H(\L)=\sum\limits_{>x,y<: x,y\in
V_n}F_{>x,y<}+ \sum\limits_{<x,y>: x,y\in
V_n}G_{<x,y>}+\sum\limits_{x\in W_n}h_i\s^z_x.
$$

The existence of the last limit easily can be checked by using
Theorem 6.2.4 \cite{BR2} (see \cite{MR}).

{\bf Lemma 3.1.} {\it Let the following condition be satisfied:
there exist integers $k_i$ and $m^{(i)}_j$, $j\in\{1,2,3\}$ and
the smallest number $\delta_i\in(0,1)$ such that $$
\frac{p_1}{p_{2}}=\delta_i^{m_1^{(i)}}, \ \
\frac{p_{11}}{p_{22}}=\delta_i^{m_2^{(i)}}, \ \
\frac{p_1}{p_{11}}=\delta_i^{m_3^{(i)}}, \ \
\exp\{h_i\}=\delta_i^{k_i}, \eqno(3.8) $$ then for the modular
group $\s^{\w_{i}}_{t}$ and the number $t_0=-2\pi/\log\delta_i$,
the equality holds $$ \s^{\w_{i}}_{t_0}=Id. $$}

{\bf Proof.} From (3.8)  we have
$$
\left. \ba{ll} p_{1}=\dsf{\delta_i^{m_1^{(i)}}}{\delta^{m_1^{(i)}}+1}, \ \
p_{2}=\dsf{1}{\delta_i^{m_1^{(i)}}+1},\\[3mm]
p_{11}=\dsf{\delta_i^{m_1^{(i)}-m_3^{(i)}}}{\delta_i^{m_1^{(i)}}+1},
\ \
p_{22}=\dsf{\delta_i^{m_1^{(i)}-m_2^{(i)}-m_3^{(i)}}}{\delta_i^{m_1^{(i)}}+1}.
\ea \right\} \eqno(3.9)
$$
Hence from (3.3),(3.4) and (3.9) we can get that
$\s^{\w_i}_{t_0}=Id$. This completes the proof.

Now using Lemma 3.1, Proposition 5.2\cite{MR} and the argument of
\cite{MR} we can prove the following

{\bf Theorem 3.3.} {\it Let $\theta_1>\sqrt{3}, \
\theta>\dsf{2\theta_1}{\theta_1^2-3}$ and the condition (3.8) be
satisfied. Then von  Neumann algebras ${\cal M}_i$ corresponding
to the translation invariant Gibbs states $\m_i$ of the Ising
model with competing interactions (1.1) on the Cayley tree $\G^2$
are factors of type III$_{\delta_i}$.}

Since $u^*_1$ and $u^*_3$ are the solution of the equation (2.6)
then from (3.8) we find that $k_1=-k_3$ and $\delta_1=\delta_3$.
This implies that the factors ${\M}_1$ and ${\M}_3$ have the same
type. It is easy to see that  $k_2=0$.

From Theorem 3.3 and the argument of \cite{MR} we have the
following

{\bf Corollary 3.4.} {\it Let $\theta_1>\sqrt{3}, \
\theta>\dsf{2\theta_1}{\theta_1^2-3}$ and the following condition
be satisfied: there exist integers $m_i$, $i\in\{1,2,3\}$ and the
smallest number $\delta\in(0,1)$ such that $$
\frac{p_1}{p_{2}}=\delta^{m_1}, \ \
\frac{p_{11}}{p_{22}}=\delta^{m_2}, \ \
\frac{p_1}{p_{11}}=\delta^{m_3}, \eqno(3.10) $$ then a von Neumann
algebras ${\cal M}_2$ corresponding to the unordered phase of the
Ising model with competing interactions (1.1) on the Cayley tree
$\G^2$ is a factor of type III$_{\delta}$. Otherwise ${\cal M}_2$
is a factor of type III$_1$.}

{\bf Corollary 3.5.} {\it Let $\theta_1>\sqrt{3}, \
\theta>\dsf{2\theta_1}{\theta_1^2-3}$ and the following conditions
be satisfied: there exist integers $k$ and $n_i$, $i\in\{1,2,3\}$
and the smallest number $\delta_1\in(0,1)$ such that $$
\frac{p_1}{p_{2}}=\delta_1^{n_1}, \ \
\frac{p_{11}}{p_{22}}=\delta_1^{n_2}, \ \
\frac{p_1}{p_{11}}=\delta_1^{n_3}, \eqno(3.11) $$ and
$$
\exp\{h_1\}=\delta_1^k, \ \ h_1>0, \eqno(3.12)
$$
then von  Neumann algebras ${\cal M}_1$ and ${\M}_3$ corresponding
to the Gibbs states $\m_1$ and $\m_3$ respectively, of the Ising
model with competing interactions (1.1) on the Cayley tree $\G^2$
are factors of type III$_{\delta_1}$. Otherwise
they are  factors of type III$_1$.}\\

{\bf Remark 3.2.} If we consider the case $\theta_1<1/\sqrt{3}$
and $\theta>\dsf{2\theta_1}{1-3\theta^2_1}$, then there are two
strictly periodic (non translation invariant) Gibbs measures. By
similar arguments as above we can prove analogical theorems as
Theorem 3.3  for these periodic measures.

{\bf Remark 3.3.} Here it would be good to mention that there is
an example of factor generated by Cayley tree, but it does not
appear from a physical system (see \cite{RR}).

It is clear that if (3.10) is not satisfied then (3.11) is too,
consequently, the algebras ${\cal M}_i$ are factors of type
III$_1$.  Suppose (3.10) is valid then (3.11) is also satisfied
with  $\delta_1\geq \delta$, more exactly, $\delta_1=\delta^r$,
where $r\in (0,1]\cap\Q$. But it is interesting whether the
equality $\delta_1=\delta$ is satisfied. The following theorem
answers to this question.

{\bf Theorem 3.6.} {\it Let $\theta_1>\sqrt{3}, \
\theta>\dsf{2\theta_1}{\theta_1^2-3}$ be satisfied. Suppose the
equalities (3.10)-(3.12) are satisfied. Then the factor ${\M}_1$
and ${\M}_3$ have types III$_{\delta^r}$, $0<r<1, r\in\Q$, while
the factor ${\M}_2$ has type III$_{\delta}$.}

{\bf Proof.} The conditions (3.10) and (3.12) imply that there
exists a rational number $s\in\Q$ such that $\t=\t_1^s$. We note
this is a necessary condition that ${\ M}_2$ to be a factor of
type III$_{\delta}$. We want to prove that $\delta_1>\delta$. Let
us assume that $\delta_1=\delta$. Keeping in mind that the numbers
$e^{2h_1}$ and $e^{-2h_1}$ are the solutions of the equation (2.6)
from (3.12) we obtain
$$
2\cosh(2k\log\delta)=\t_1^2-2\t_1^{1-s}-1 \eqno(3.13)
$$
here we have used that $\a=2\t_1^{1-s}-\t_1^2$. From (3.11) we
find that $\log\delta=-2n_1J_1\b$, substituting it into (3.13) we
have
$$
2\cosh(4nJ_1\b)=\t_1^2-2\t_1^{1-s}-1, \eqno(3.14)
$$
here without loss of generality we may assume that $n>0$, $n\in
\Z$, since $\cosh(x)$ is an even function. Defining
$f(n)=2\cosh(4nJ_1\b)$ from (3.14) it is easy to see that
$f(1)>\t_1^2-2\t_1^{1-s}-1$, since $\t_1>\sqrt{3}$. It is clear
that $f(n)$ is an increasing function, so this implies that the
equality (3.14) can not be satisfied  for any positive integer
$n$. This means $\delta_1>\delta$. Consequently, the factors
${\M}_1$ and ${\M}_3$ can not have the same  type with the factor
${\M}_2$. This completes the proof.

The proved Theorem means that the factor ${\M}_2$ can be
considered as a subfactor of ${\M}_1$ and ${\M}_3$ respectively.

{\bf Corollary 3.7.} {\it Let $\theta_1>\sqrt{3}, \
\theta>\dsf{2\theta_1}{\theta_1^2-3}$ be satisfied. If there is an
irrational $\gamma$ such that $J=\gamma J_1$ then the factors
${\M}_i$ ($i=1,2,3$) have type III$_1$. }

Let us consider some more concrete examples of factors.

{\bf Example 3.1.} Suppose that $J=0$ and $\t_1>\sqrt{3}$. Then
the condition $\t>\dsf{2\t_1}{\t_1^2-3}$ implies that $\t_1>3$. In
this case the equality (3.10) reduces to the following one
$$
\dsf{p_{11}}{p_{22}}=\delta^m,
$$
here as before $\delta\in (0,1)$ and $m\in\Z$, which is automatically satisfied with
$\delta=\t_1^{-1}$ and $m=-1$. So in this case ${\cal M}_2$ is a factor of type III$_{\delta}$.
But it is interesting question is whether  the factors ${\cal M}_1$ and ${\cal M}_3$
can have type III$_{\delta_1}$, while the
factor ${\cal M}_2$ has type III$_{\delta}$. Now we going to show that this can be occur.

Indeed, we firstly note that in the considered case the equation
(2.4)  can be written as follows
$$
h=2arctanh(\tilde\t\tanh h) \eqno(3.15)
$$
here $\tilde\t=\tanh(J_1\b)$ (see, \cite{BRZ1}).
We recall that the condition $\t_1>3$ is equivalent to
$\tilde\t>\dsf{1}{2}$.  Now using the formula
$$
\tanh(2x)=\dsf{2\tanh x}{1+\tanh^2{x}}
$$
from (3.15) we obtain
$$
\tanh h_1=\dsf{2\tilde\t\tanh h_1}{1+(\tilde\t\tanh h_1)^2},
$$
it yields that
$$
h_1=arctanh\bigg(\dsf{\sqrt{2\tilde\t-1}}{\tilde\t}\bigg).
\eqno(3.16)
$$
Let us turn to the conditions (3.11) and (3.12). In our case they
can be reduced to the following ones
$$
\dsf{p_{11}}{p_{22}}=\delta_1^n, \ \ \ \exp\{h_1\}=\delta_1^k, \
n,k\in\Z. \eqno(3.17)
$$

Choose the number $\tilde\t$ such that which satisfies the following equation
$$
\tilde\t^3+5\tilde\t^2+7\tilde\t-5=0. \eqno(3.18)
$$
It is not hard to check that the required $\tilde\t$ does exist,
i.e. with the property $1/2<\tilde\t<1$. Put
$\delta_1=\sqrt[4]{\delta}$ or $\delta_1=\t_1^{-1/4}$. It easy to
see that for such $\delta_1$ we have $n=-4$.  From (3.17) we find
$$
h_1=-\dsf{k}{2}J_1\b,
$$
which yields
$$
h_1=-\dsf{k}{2}arctanh\tilde\t, \eqno(3.19)
$$
here we have used the $J_1\b=arctanh\tilde\t$.

We will show that this equality is satisfied when  $k=-1$. Indeed,
using (3.16) from (3.19) we have
$$
\dsf{\sqrt{2\tilde\t-1}}{\tilde\t}=\tanh\bigg(\dsf{
arctanh(\tilde\t)}{2}\bigg). \eqno(3.20)
$$
Now according to the formula
$$
\tanh\dsf{x}{2}=\dsf{1-\sqrt{1-\tanh^2x}}{\tanh x}, \ \ x>0
$$
from (3.20) we get
$$
\sqrt{2\tilde\t-1}+\sqrt{1-\tilde\t^2}=1.
$$
The last equation equivalent to the following one
$$
(\tilde\t-1)(\tilde\t^3+5\tilde\t^2+7\tilde\t-5)=0.
$$
The condition (3.18) yields that the last equality  is satisfied,
hence (3.20) is valid. Thus the factors ${\cal M}_1$ and ${\cal
M}_3$ have type III$_{\sqrt[4]{\delta}}$.

These results clarifies and specifies the results obtained in
\cite{M},\cite{MR}.

{\bf Example 3.2.} Suppose that $J=J_1$ and $J\neq 0$, this means
$\t=\t_1$. Hence the equality (3.10) is satisfied with parameters:
$\delta=\t^{-1}$, $m_1=-1$, $m_2=-1$, $m_3=0$. So according to
Corollary 3.4 we conclude that ${\cal M}_2$ is a factor of type
III$_{\delta}$. Now assume that there is a phase transition, i.e.
the condition $\t_1>\sqrt{3}$, $\t>\dsf{2\t_1}{\t_1^2-3}$ is
satisfied, which implies in our case (i.e. $J=J_1$) that
$\t>\sqrt{5}$. Now we are going to find another $\delta_1$ for
which the factors ${\M}_1$ and ${\M}_3$ have type
III$_{\delta_1}$.

Put $\t=1+\sqrt{2}$ and $\delta_1=\sqrt{\delta}$. It is clear that
(3.11) is satisfied. Now we should check (3.12). Keeping in mind
that the numbers $e^{2h_1}$ and $e^{-2h_1}$ are the solutions of
(2.6) from (3.12) we get
$$
\delta^k+\delta^{-k}=\t^2-3. \eqno(3.21)
$$
Put $k=1$. Let us show this equality is satisfied. The equality
(3.21) can be written as follows
$$
(\t+1)(\t^2-2\t-1)=0.
$$
The chosen $\t$ satisfies this equation, hence (3.12) is valid.
This is the required.

If $J=0$ then the phase transition does not occur and the factor
${\cal M}_2$ has type II$_1$. We note in the case $\t_1>\sqrt{3}$
and $\t>\dsf{2\t_1}{\t_1^2-3}$ the factor  ${\cal M}_2$ can not
have a type II$_1$.\\

\section{Discussion of the results}

It is known that to exact calculations in statistical mechanics
are paid attention by many of researchers, because those are
important not only for their own interest but also for some deeper
understanding of the critical properties of spin systems which are
not obtained form approximations. So, those are very useful for
testing the credibility and efficiency of any new method or
approximation before it is applied to more complicated spin
systems.  In the previous paper \cite{MR} we have exactly solved
an Ising model on a Cayley tree, the Hamiltonian of which
contained ternary interactions. In addition, we found some
conditions on parameters which enabled to determine exactly  types
of von Neuamann algebras associated with periodic Gibbs states of
that model. In the present paper we continue investigations of the
Ising model, but now we consider a model with the
next-nearest-neighbor binary interactions. Using the same way as
\cite{MR} we exactly solve a phase transition problem for the
model, namely,  we calculated critical curve such that there is a
phase transitions above it, and a single Gibbs state is found
elsewhere. Comparing with the results of \cite{MR} in the present
paper we describe all periodic Gibbs states associated with
subgroups of $G_2$ with finite index, while in the mentioned paper
we only found $G^{(2)}_2$-periodic Gibbs states. Besides, we also
find ground states of the considered model. Here (in the paper) as
in \cite{MR} we also find some conditions of parameters $J$ and
$J_1$ which completely determine types of von Neumann algebras
corresponding to the translation-invariant Gibbs states, but now
we show how these algebras related with each other, more precisely
speaking, we prove that the factor corresponding to the unordered
phase is a sub-factor of the factors associated with the minimum
and maximum Gibbs states. We note that this kind of question was
not considered in \cite{MR}. Finally, we demonstrate some more
concrete examples of such factors, which clarify the results
obtained in \cite{M},\cite{MR}.

We note that some computer simulations  results of the model
considered were studied in \cite{MTA}. Some other phase
transitions problems were considered  in \cite{L}.

{\bf Acknowledgements} The work was done within the scheme of
Borsa di Studio CNR-NATO. One of authors (F.M.) thanks CNR for
providing financial support and II Universita di Roma "Tor
Vergata" for all facilities. Besides, he also thanks
Prof.E.Presutti for kind hospitality and useful discussions. U.R.
thanks Institute des Hautes Etudes Scientifiques (IHES) for
supporting the visit to Bures-sur-Yvette (IHES, France) in
September-December 2003. The work is also partially supported by
Grant $\Phi$-1.1.2 Rep. Uzb.

 \edc